# Evidence of hydrogen termination at grain boundaries in ultrananocrystalline diamond/hydrogenated amorphous carbon composite thin films synthesized via coaxial arc plasma

Naofumi Nishikawa[1,2,*,†]

[1] *Department of Applied Science for Electronics & Materials, Kyushu University, 6-1 Kasuga, Fukuoka 816-8580, Japan*
[2] *School of Materials Science, Japan Advanced Institute of Science and Technology (JAIST), 1-1 Asahidai, Nomi, Ishikawa 923-1292, Japan*
[*] *Author to whom correspondence should be addressed: naofumi_nishikawa@kyudai.jp*
[†] *Present: Independent Researcher*

(Dated: December 13, 2020)

Ultrananocrystalline diamond/hydrogenated amorphous carbon composite thin films consist of three different components; ultrananocrystalline diamond crystallites, hydrogenated amorphous carbon, and grain boundaries between them. Since grain boundaries contain a lot of dangling bonds and unsaturated bonds, they would be a cause of carrier trap center degrading device performance in possible applications such as UV photo-detectors. We experimentally demonstrate hydrogen atoms preferentially incorporate at grain boundaries and terminate dangling bonds by means of several spectroscopic techniques. XPS measurements cannot detect quantitative transitions of $sp^2$- and $sp^3$-hybridized carbons in the films, resulting in 55–59% of $sp^3$ contents. On the other hand, FT–IR and NEXAFS exhibit some variations of the amounts of certain carbon hybridization for sure. The former confirms the transformation from $sp^2$ to $sp^3$ hydrocarbons by ~10% by additional hydrogenation, and the latter represents chemical configuration changes from $\pi^*$ C≡C and $\pi^*$ C=C to $\sigma^*$ C−H as well as more $\sigma^*$ C−C contained. These results can be an evidence of localized hydrogen at grain boundaries, which plays a part in terminating dangling bonds and unsaturated bonds, and they are correlated with the optical and electrical properties of the films investigated in some previous research. Our spectroscopic studies on the hydrogenation effects combined with the discussion on the optical and electrical characteristics confirm that the hydrogenation can be an effective tool of an enhancement of photovoltaic performance in the above sensing applications.

## I. INTRODUCTION

Diamond is well known for its tetrahedral structure and configuration of $sp^3$-hybridized carbon, giving rise to some original characteristics; for example, its super hardness, excellent thermal conductivity, chemical inertness, and electrically intrinsic in case of undoped. Diamond is able to be synthesized artificially by means of a variety of chemical vapor deposition (CVD) techniques in common, such as hot-filament CVD (HFCVD),[1,2] microwave plasma CVD (MWCVD),[3] and radio-frequency plasma CVD (RFCVD),[4,5] which perform depositions mainly under methane and hydrogen and/or argon gases atmosphere. These man-made diamonds are divided based on their crystallinity into single-crystal diamonds or poly-crystalline diamonds. Furthermore, the latter can be classified in more detail according to their grain size; micro-crystalline diamonds (MCDs) or nano-crystalline diamonds (NCDs) in 10 to 100 nm and a few hundred nanometers to several hundred micrometers in diameter, respectively. Especially, NCDs comprising crystallites with less than 10 nm in grain size are sometimes called ultra-nano-crystalline diamonds (UNCDs).[6] Single-crystal diamonds are composed of highly ordered carbon atoms, hence they possess long-range order in atomic scale, on the contrary, poly-crystalline diamonds consist of carbon atoms oriented totally at random, and of large number of interfaces, namely grain boundaries (GBs). In addition, meta-stable fractions are inevitably contained in poly-crystalline diamonds, which are different from diamond crystallites, and they are called amorphous carbon (a-C) or hydrogenated-amorphous carbon (a-C:H).[7] In general, amorphous fractions tend to expand and the number of GBs among diamond grains increase with the decrease of the grain size, therefore, the properties of a-C and a-C:H are reflected more conspicuously on those of the whole diamonds. Since a-C and a-C:H fractions don't possess certain crystallinities, it is difficult to describe the physical properties precisely. In spite of it, there have been some theoretical studies predicting some of them such as electrical,[8,9] structural[9–13] ones over the past few decades, and recent research has realized their low-dimensional morphologies experimentally.[14,15] In addition, intriguing multi-functional carbon materials engineered from an a-C named quenching-carbons (Q-carbons) that exhibit extreme hardness,[16] superconductivity by heavy boron-doping[17–19] have appeared and been explored for the past few years. On the other hand, general NCDs and UNCDs are rather composite materials with such an a-C and/or a a-C:H, and therefore, they necessarily contain GBs that include various chemical bonding configurations and dangling bonds, which makes the problem more tough and complicated. That's the reason why it is worthy of much attention from scientists and vigorous investigations and discussion have been done both from theoretical and experimental approaches in recent years.

As stated above, while UNCD thin films are grown under argon-rich atmosphere preventing an enlargement of diamond grains and fostering re-nucleation via CVD methods,[6] Yoshitake's group has realized UNCD fabrications by use of two other techniques. The one is pulsed laser deposition (PLD) method that can prepare the films by employing excimer laser ArF as an excitation light source, and irradiating it on a solid graphite target under hydrogen atmosphere.[20] Another is coaxial arc plasma deposition (CAPD) technique, which can synthesize the films by applying pulsed voltages directly to a graphite target.[21] In each UNCD fabrication process, CVDs utilize seed crystals as a precursor and repeat nucleation to prepare the films with dense UNCD crystallites. On the other hand, physical vapor deposition (PVD) methods such as PLD and CAPD perform nucleation in each pulse of excimer laser irradiation or arc discharge, resulting in the forms of UNCD grains being embedded into a-C or a-C:H matrices and interspersed inside them.[22,23] As a result of it, UNCD films with higher compositeness are synthesized, which can also be called UNCD/a-C or UNCD/a-C:H composite films and also be written as "UNCD/a-C" or "UNCD/a-C:H" films as firstly named by C. Popov et al.[24] a-C and a-C:H that belong to a family of diamond-like carbons (DLCs), consist of a variety of $sp^3$- and $sp^2$-hybridized carbons, and the ratio of $sp^3$ to $sp^2$ carbons have a large effect on the physical properties of UNCD/a-C and UNCD/a-C:H films. Particularly, CAPD method is the most functional in controlling hydrogen content in the films among all techniques mentioned above, since it's able to grow the films with no hydrogen gas.[25] Hence, it is vital for predictions on many sorts of features that the films by CAPD possess to investigate how and how to extent hydrogenation exerts influences on their chemical bonding configurations, i.e. $sp^3$ and $sp^2$ bonds, which largely affect the physical properties of the films.

UNCD/a-C and UNCD/a-C:H films synthesized by CAPD have been studied up to the present, by taking the advantages of conventional diamonds, toward the applications not only to hard coatings[26–30] and thermoelectric conversion elements,[31] but also to optoelectronic devices such as photovoltaics.[32–44] As mentioned above, UNCD/a-C:H films prepared by CAPD include more a-C and/or a-C:H fractions and GBs that would act as generation sources of photo-induced carriers, which might be a cause of extremely large light absorption coefficients excessing $10^5$ cm$^{-1}$ in the photon energy range of 3 to 6 eV.[23] This value is at least two orders of magnitude larger than those of the films synthesized by common CVD techniques, that's why UNCD/a-C:H films by CAPD are preferable potential candidates for UV photo-detector applications.

Actually to date, p- and n-type conductions of the UNCD/a-C:H films by CAPD have been already realized by doping boron[32,33] and nitrogen,[36] respectively in experimental conditions, and p-n heterojunction diodes have been also fabricated by depositing these films on n-[32] and p-type[36] single-crystalline Si substrates. In recent research, it has been clarified that hydrogenation would be a key factor of an enhancement of photovoltaic performance in these diodes.[34,35,41] And as a result of the latest study, it was indirectly confirmed through electrical properties of the diodes that hydrogenation would be assisted with a control of applied pulsed voltages to a graphite target equipped with an arc gun, by customizing an apparatus similar to CAPD one in principle.[41] However, it is unclear how hydrogenation induced with this technique has an effect on the chemical bonding states of the films.

In this work, we characterize the UNCD/a-C:H films by X-ray photoelectron spectroscopy (XPS), Fourier transform infrared spectroscopy (FT–IR), and near-edge X-ray absorption fine structure (NEXAFS), to investigate hydrogenation effects on chemical bonding configurations of the films. From the experimental results, we demonstrate quantitative transitions of specific chemical bonds, which can be an evidence of hydrogenation localized at GBs of the films. Furthermore, these spectroscopic results are correlated with the electrical and optical characteristics of the films, which are highly relevant to the defects at GBs and in a-C and/or a-C:H matrices, reported until today, in order to bring about some informative discussion on the applications to optoelectronic devices at every important point. Interfacial phenomena that are attributed largely to such regions and often exhibit anomalous effects different from those occur in a bulk are hot issues at present and will be interests from the point of view of physical science, and their brief perspectives are provided in a final section.

## II. EXPERIMENTAL

UNCD/a-C:H films doped with a small amount of boron were prepared on insulating single-crystalline Si substrates by changing an applied voltage to an arc gun with CAPD method. After the Si wafer was cut into substrates with their dimensions of 15 × 15 mm, they were dealt with clean water, ethanol, and acetone solutions by ultrasonic cleaning for five minutes in series. The processed substrate was treated with HF solutions to remove oxidized fractions inside itself, and it was equipped with a sample holder in a vacuum chamber soon afterwards. Boron-doping in the films will be realized by operating arc discharge by use of a graphite target blended with 1.0 atomic (at.) % boron content, equipped with an arc gun, as similar to a previous study.[32] Inside of a chamber was evacuated down by a turbo molecular pump adequately to induce plasma, being kept a hydrogen pressure of 53.3 Pa. The arc discharge is carried out by adding pulsed voltage to the cathode attached with the target, resulting in the occurrence of plasma constituting atomized, ionized, dimerized carbon and boron to deposit accompanied by hydrogen atoms onto the substrate to form boron-doped UNCD/a-C:H films with thickness of ~200 to 300 nm. The detailed experimental parameters are summarized in Table I.

Since several studies have reported that hydrogen-terminated diamond surfaces exhibit week p-type conductions in some kinds of CVD diamonds,[45–47] it is considered to be preferable for consistent discussion being conducted to prepare the films mentioned above, based on a hypothesis that our films also possess the same or similar conductions. These films were deposited with applied voltages of 70, 85, 100 and 115V to the arc gun and were synthesized under the same preparation conditions with our recent paper,[41] which can be discussed comparatively correlating with some electrical characteristics. When the applied voltage increases, the amount of particles ejected from the target also increases. On the other hand, hydrogen pressure in a chamber is constant and therefore the relative hydrogen content of which is involved with ejected carbon and boron particulates and of the films prepared in actual decreases contrarily. Thus hydrogen content in the films would be controllable by changing an applied pulsed voltage, as indicated in a recent paper.[41] The synthesized films would consist of UNCD crystallites with estimated grain size of 2–3 nm considering some experimental results similarly set up that report the grain diameter of 1.6–2.3 nm for un-doped films[23,33] and the boron-doping enlarges the grain size up to ~11.1 nm for the films fabricated by using a target with 5.0 at. % boron content[33] while the influence would be limited due to the slightest boron-doping in our case. Regarding the latest advancement of the boron-doping mechanism and the effect on chemical bonding structure can be referred to Ref. 48. Well-mapped atomic force microscope (AFM) images of UNCD/a-C:H films fabricated via CAPD method can be found in Refs. 43 and 49, which will be helpful in facilitating comprehension of the film structures. Although transmission electron spectroscopy (TEM) image reported only in case of the films prepared by PLD can be found in Ref. 20 which illustrates film/substrate structures, that technique, being employed to the films consisting of composite structures with UNCDs, has some obstacles in capturing clear pictures, for instance, the UNCD particles are disintegrated by irradiation unless focused ion beam (FIB) milling process is carried out. XPS and NEXAFS measurements were performed at beamline 12 of Kyushu Synchrotron Light Research Center. Mg Kα line ($h\nu$ = 1253.6 eV) and 350 eV X-rays were employed as light sources of both measurements. FT–IR spectra were obtained by FT/IR–4200 type A of DJK Corporation.

**Table I.** Experimental setup parameters when UNCD/a-C:H films were prepared by utilizing CAPD system. All those presented below are concerning a fraction in CAPD system, i.e. one part of CAPD apparatus or its adjunct.

| Base pressure (Pa) | Substrate temperature (°C) | Frequency of pulse (Hz) | Distance between target head and substrate (mm) | Capacitance (μF) |
|---|---|---|---|---|
| $< 5.0 \times 10^{-5}$ | 550 | 5 | 15 | 720 |

### III. RESULTS AND DISCUSSION

#### A. XPS characterization

Figure 1 shows the wide-scan XPS spectra in boron-doped UNCD/a-C:H films with different applied voltages to the arc gun. The most prominent peaks in every spectrum in the photon range of 285 ± 2 eV indicate an existence of carbon 1s (C 1s) orbitals. This means that the largest constituents of the films are carbon-system materials such as diamonds, a-Cs, and a-C:Hs. The secondly distinct peaks positioned at 532 ± 3 eV are equivalent to oxygen 1s (O 1s) spectra, which will be attributed to surface oxidation of the films, since argon-ion bombardment treatments would clearly vanish them as reported previously.[50] Peak intensities at 191 ± 1 eV originating from boron 1s (B 1s) orbitals are too weak to find as compared to those of C 1s and O 1s ones. However, it's natural signal that the B 1s intensity is weak because the atomic sensitivity factor or the cross section of the B 1s which affect the peak intensity in XPS is small compared to the C 1s and the O 1s, as the similar observations have been confirmed in the films containing more boron and fabricated by CAPD method similarly.[34,48] The magnified fraction that emphasizes the transition from B 1s orbital to $\pi^*$ states is presented in Fig. 1(b). According to a recent study,[44] the estimated density of boron atom in UNCD/a-C:H films fabricated using a graphite target with its amount of 1.0 at. % via CAPD is around 0.5 at. %, and this value would be reliable since the preparation conditions and other experimental parameters are almost same with ours. The spikes of the peaks located at around 975 eV would be attributable to Carbon KVV Auger transitions, as the similar jumps were observed in both boron-doped MCD[51] and a-C:H[52] thin films, all of which are summarized and illustrated in Fig. 1(c). The wide-scan XPS spectra of our samples have the same line shapes as previous studies about the boron-doped UNCD/a-C:H films.

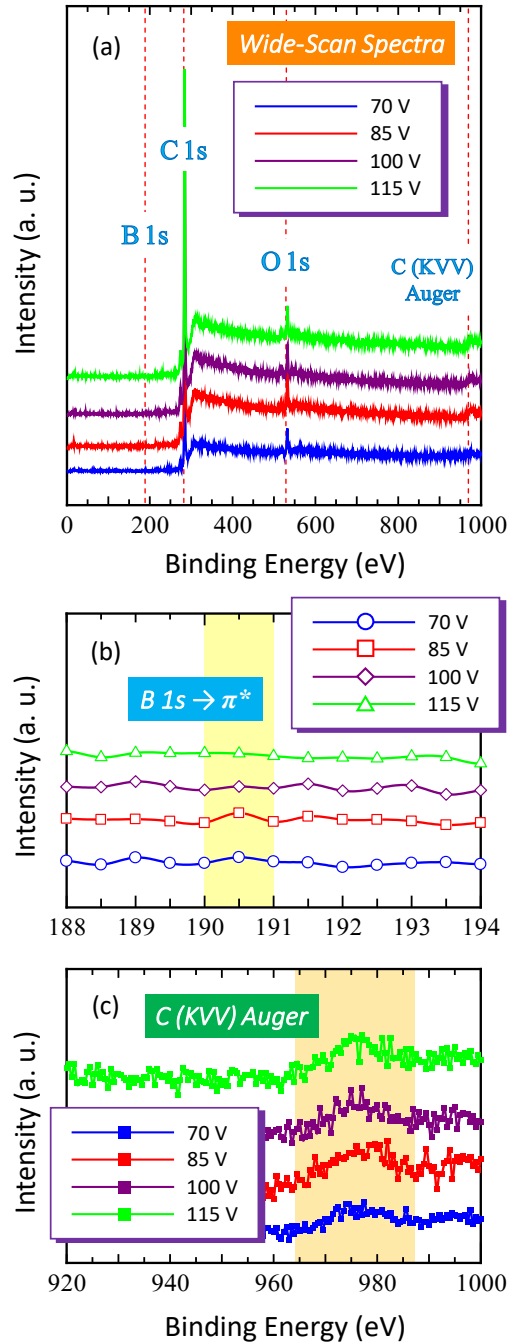

**Fig 1.** (a) Wide-scan XPS spectra of boron-doped UNCD/a-C:H films deposited via coaxial arc plasma, with applied voltages to a coaxial arc gun of 70, 85, 100 and 115V. (b) and (c) are magnified portions of the wide-scan to represent clearly the B 1s and C(KVV) Auger transitions. Soft X-ray (Mg Kα, with a wavelength of 1253.6 eV) was employed to the four samples as a light source.

To be discussed quantitatively about the carbon hybridizations comprising the films, C 1s spectra were examined for an assessment of $sp^3$-bonded carbons relative to $sp^2$-bonded ones. After the charge-up effects were calibrated empirically based on the database of UNCD/a-C:H films measured at a BL 12 accumulated up to the present, all the spectra were normalized to make them possible to compare with each other. Normalized spectra are illustrated in Fig. 2(a) with the original ones in the inset. All of them seem to be almost overlapped mutually. The reason is discussed afterwards accompanied with the next results of the deconvoluted spectra. To investigate variations of the relative amounts of $sp^2$- and $sp^3$-coordinated carbons contained in the films by hydrogenation, these spectra were deconvoluted by use of a fitting program 'XPS peak 41'. All of the spectra are deconvoluted by Voigt curves consisting of Gaussian and Lorentzian functions, after the backgrounds are subtracted by Shirley's method.[53,54] The references of the peak positions describing $sp^2$- and $sp^3$-hybridized carbons and C−O−X at 284.5, 285, 286 eV, in order are employed to decompose the spectra, which successfully deconvoluted the spectra of the films synthesized under the most similar conditions by CAPD in a recent work.[31] As shown in Figs. 2(b)–(e), the full width at half maximum (FWHM) of the spectra of $sp^3$-coordinated carbon falls within the range of 0.97 to 0.99 eV. These values are almost the same with those of UNCD/a-C (0.99 eV)[21] and UNCD/a-C:H (0.98 eV)[23,26] films fabricated by CAPD as well as a-C (1.04 eV)[22] and a-C:H (1.06 eV)[22] films by PLD reported previously. These values would be within error range in the peak fitting process since the FWHM of the components in XPS spectra is very sensitive to the equipment condition and the peak fitting process such as the background subtraction and the relative peak positions, and further discussion on it is not done. The values of $sp^3/(sp^2+sp^3)$ range from 55 to 59%, which are situated between the values of 64 and 41% in case of UNCD/a-C:H and UNCD/a-C films by CAPD, reported recently.[31] This is in good agreement with a notion of hydrogenation transforming $sp^2$ bonds at GBs and an a-C and/or a a-C:H fractions to $sp^3$ bonds, resulting in the increase of the amount of $sp^3$ bonds relatively. On the other hand, the hydrogenation effect scarcely changes both the values of FWHMs of $sp^3$ and $sp^3/(sp^2+sp^3)$ ratio. This might indicate that XPS cannot detect the hydrogen-terminated carbons as a previous work explained in case of a-C:H films by RFDVD,[55] taking into consideration that the hydrogenation certainly enhanced photo-induced carrier under UV illumination, of the diodes comprising these UNCD/a-C:H films fabricated under totally the same conditions, which implies the hydrogenation would affect the morphology of the films.[41] Or hydrogen might be appended to oxygen at the film surface to form hydroxyl carbon that will also not to be detected based on a previous report as for boron-doped MCD.[56] Both of them are attributable to the limitation of XPS resolution. Hence, our UNCD/a-C:H films are not sensitive to XPS analysis, which reflects well on the almost identical pictures of photoelectron C 1s spectra as shown in Fig. 2(a), in spite of a control of the applied voltages that showed vivid variations of electrical characteristics.[41] In the meantime, some of the previous works certainly detected the relative increase of $sp^3$-bonded carbons by means of hydrogenation that were conducted by changing base pressure of hydrogen in a chamber.[26,31] The reason for it is unclear at present. The most reasonable interpretation here is the hydrogenation by controlling the applied voltage might have occurred inside the films enough deeply not to be detected by X-ray photoelectron, different from those by increasing hydrogen pressure in a chamber, because X-ray can penetrate only several nanometers at the film surface to investigate chemical bonging configurations within this range.

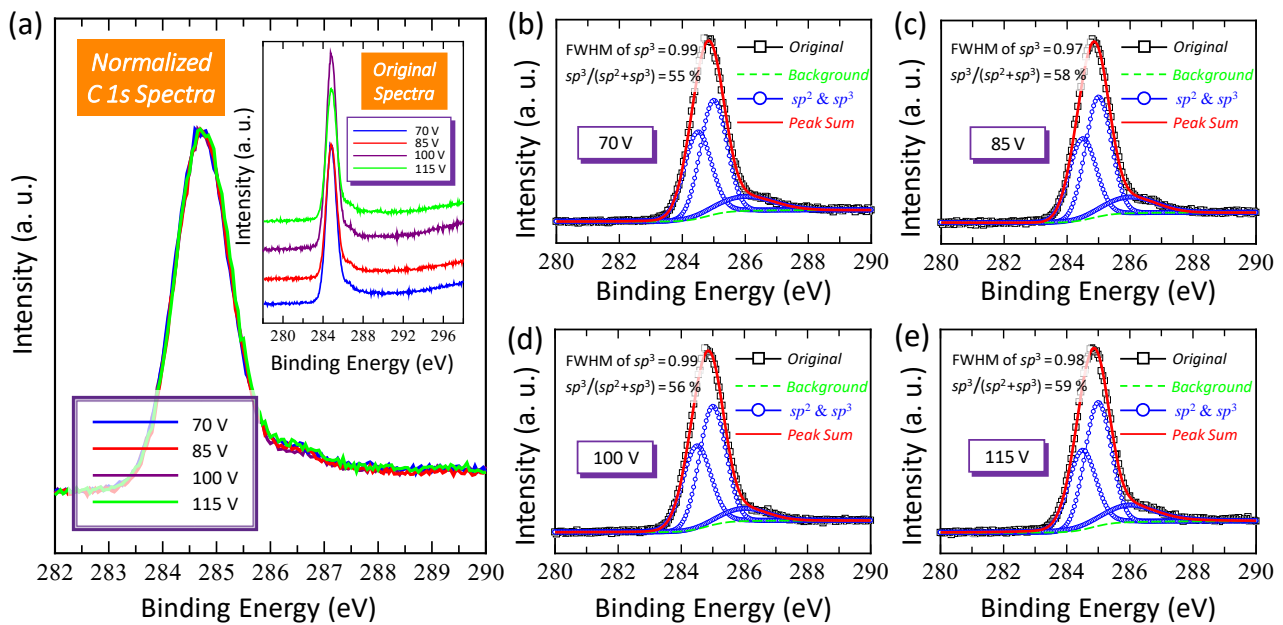

**Fig. 2.** (a) Normalized carbon 1s (C 1s) XPS spectra in boron-doped UNCD/a-C:H films with their original spectra in the inset. (b)–(e) illustrate each spectrum separately, accompanied with their backgrounds (illustrated with light green curves), deconvoluted spectra (originating from $sp^2$-, $sp^3$-bonded carbons, and carbonyl group positioned at the binding energies of 284.5, 285 and 286 eV, respectively) (blue), and the sum of them (red).

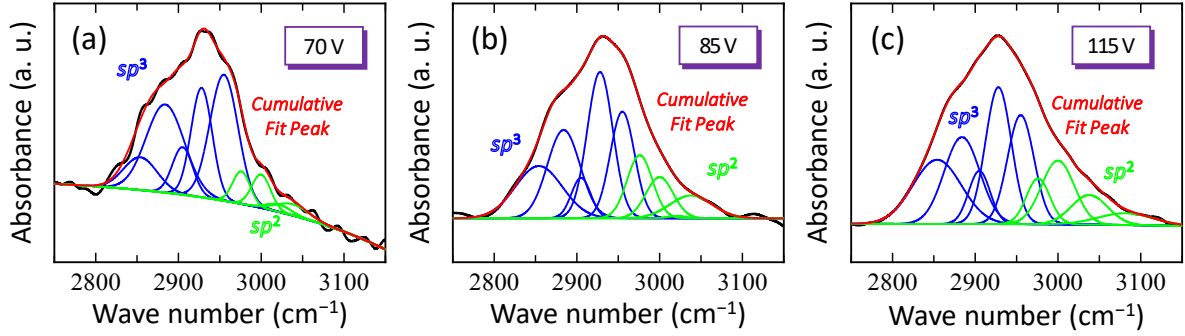

**Fig. 3.** FT–IR spectra of boron-doped UNCD/a-C:H films fabricated with applied voltages of (a) 70, (b) 85, and (c) 115 to an arc gun. Baselines were set with proper approximations to remove an effect of interference fringes. All the peak deconvolutions were performed based on the previous report[22] as reference with gaussian functions. Concretely, each peak position is centered at 2854 eV (Symmetric $sp^3$-$CH_2$), 2884 eV (Symmetric $sp^3$-$CH_3$), 2905 eV ($sp^3$-CH), 2928 eV (Asymmetric $sp^3$-$CH_2$), 2955 eV (Asymmetric $sp^3$-$CH_3$), 2976 eV (Olefinic $sp^2$-$CH_2$), 3000 eV (Olefinic $sp^2$-CH), 3021 and 3038 eV (Aromatic $sp^2$-CH), and 3082 eV (Asymmetric $sp^2$-$CH_2$) of wave number (cm$^{-1}$). Peak curves originating from $sp^3$- and $sp^2$-coodinated hydrocarbons are colored blue and light green, respectively, and the total sum of them are illustrated with red ones.

### B. FT–IR characterization

FT–IR was employed to obtain C−H stretching bands in the region between 2800 and 3100 cm$^{-1}$, which is functional in quantitatively determining the chemical bonds of some hydrocarbons in the UNCD/a-C:H films, with the variation of hydrogen contents that would be changed with the control of an applied voltage to an arc gun. The results are illustrated in Fig. 3 except for the case of 100 V which represented its atypical form. All the FT–IR spectra are well deconvoluted by gaussian functions after baselines were set to be approximated properly. This time the peak positions in all decomposed spectra stemming from $CH_n$ ($n$ = 2, 3) bonds are referred to a previous work[22] that could've deconvoluted spectra well obtained in some subsequent works. Especially for a case that an applied voltage is 70 V [Fig. 3 (a)], prominent interference fringe effect is observed, hence it is removed by approximating baseline from the whole figure.

An infrared (IR) absorbance of the films prepared with an applied voltage of 115 V was obviously larger than those of the films in 70 and 85 V, which exhibited almost the same intensities of absorbance. This might be an implication that less GBs, which would bestow a salient feature of an extremely large light absorption coefficient to the films, are contained in the former compared to the latter. As reported by a couple of literatures that examined the electron spin resonance (ESR) spectroscopy, the density of dangling bonds included at GBs is an order of 10$^{22}$ to 10$^{23}$ cm$^{-3}$ in case of the films by 100 V,[43,57] hence the estimated density is larger than this value in those of 70 and 85 V, and vice versa. The grain size of UNCD crystallites in the undoped UNCD/a-C:H films synthesized via CAPD is inclined to be enlarged from 1.6 nm up to 2.3 nm by hydrogenation, based on the latest technological advancements,[25,26] and it also be enlarged by doping boron,[49] which provably leads to the estimated grain size of our films equivalent to ~2.3 nm or larger insignificantly. Taking into account these morphological natures revealed until today, it's natural to consider that the abundant IR absorbance in higher applied voltage suggests the less existence of UNCD grains with its almost identical grain size, as well as less GBs as described above.

As a result of the peak deconvolution, symmetric $sp^3$-$CH_3$ (2884 cm$^{-1}$) and asymmetric $sp^3$-$CH_3$ (2955 cm$^{-1}$) are confirmed to be increased with less applied voltages. In addition, asymmetric $sp^2$-$CH_2$ (3082 cm$^{-1}$) are clearly vanished in accordance with that. The reason for it would be dangling bonds at GBs between UNCD grains and a-C and/or a-C:H fractions that are composed of $sp^2$-$CH_2$ are terminated by hydrogenation and converted to $sp^3$-$CH_3$. This interpretation is in consistency with the previous results of research on the photovoltaic actions in the diodes consisting of the boron-doped UNCD/a-C:H films fabricated under completely the same conditions.[41] Regarding $sp^2$-bonded hydrocarbons as they colored light green in Fig. 3, there are all sorts of them state of existence in any cases. This would indicate an incorporation of typical a-C and a-C:H since a common knowledge of a-C and a-C:H films describes that a variety of $sp^2$- and $sp^3$-hybridized hydrocarbons are contained in them and the amount of $sp^1$-bond is negligibly small compared with those of them.[58] Therefore an a-C:H fraction surely occupies boron-doped UNCD/a-C:H thin films to a certain extent as previously studied in case of undoped films.[22,23] The ratio of the total amounts of $sp^2$-hybridized hydrocarbons consisting the films against those of all sorts of hydrocarbons changes from ~23% (85 and 115 V) to 13% (70 V), by reducing an applied voltage, i.e. increasing presumable hydrogen content in the films. As we can easily confirm from the deconvoluted spectra [blue and light green colored in Fig. 3], the only $sp^2$- and $sp^3$-bonded hydrocarbons consist of the entire films in all cases. Hence, it is considered that a portion of value differences calculated at 10% $sp^2$-bonded hydrocarbons is terminated by additional hydrogen and deformed to $sp^3$-bonded ones. These value transitions are in good agreement with the results of current-voltage (I–V) characteristics of the heterojunction diodes (compare figures in Ref. 41), that's why all of the results and discussion in this subsection explains quite well the previous experimental results.

### C. NEXAFS characterization

NEXAFS technique, which is a powerful tool of local bonding configuration analysis of both organic and inorganic materials,[59] was employed to investigate those of our films in more detail. NEXAFS spectra of the films deposited with applied voltages of 70, 85, 100 and 115 V are shown in Fig. 4.

The NEXAFS spectra of our films would be superposition of wave functions of electrons excited both from graphitic bonds contained in a-C and/or a-C:H, GBs and covalent bonds in UNCD grains, considering their components. To facilitate discussion, every spectrum shall be divided into two components at a photon energy of ~291 eV where a large absorption begins, based on their contributions. A portion ranging no more than the dividing point derives from a-C and/or a-C:H,[61,62] and the other does

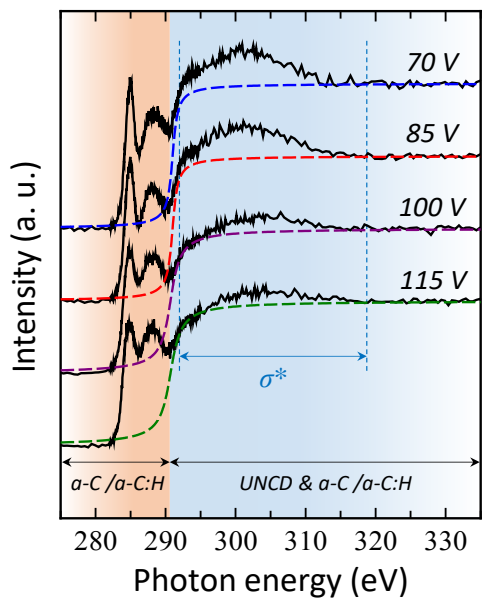

**Fig. 4.** NEXAFS C K-edge spectra, measured by total fluorescence yield (TFY) mode, of the films synthesized with applied voltages of 70, 85, 100 and 115 V. Pre-edge ranging a photon energy of no more than 283 eV and post-edge more than 314 eV are normalized as similar to a previous study.[60] Each dashed curve expresses $\sigma^*$ excitation related to diamond approximately fitted with arctangent function steps.

UNCD grains as well as an amorphous part. The former is attributed mainly to $\pi^*$ transition, however $\sigma^*$ transition is also the cause of it. This fraction is highly convoluted with some spectra originating from different bonding excitations and the details are discussed afterwards. The latter, comparatively broad bands originates from several $\sigma^*$ resonances. In order to elucidate these contributions, arctangent steps (dashed curves) are fitted to the spectra corresponding to an onset of $\sigma^*$ excitation after the post-edges are normalized. Although these broad $\sigma^*$ bands are due to both UNCDs and an a-C, a a-C:H, a sharp core exciton peak positioned at around 289.5 eV, just before an outset of the $\sigma^*$ absorption and a dip stemming from second band gap at 302.5 eV,[63,64] both of which are original to diamonds, have not been confirmed. This might be owing to a structural feature in UNCD/a-C:H films synthesized by CAPD, most of which comprise an amorphous part and the amount of diamonds are not enough large to be detected clearly by NEXAFS. In addition, it might be a supportive explanation that there have been a couple of reports that UNCDs exhibit less noticeable core exciton edge of diamond compared with that in single-crystal diamonds.[65,66] It can be confirmed that the intensity of this $\sigma^*$ band increases with diminishing an applied voltage. This might be because hydrogenation induced by controlling an applied voltage stabilizes diamond surface and/or an amorphous fraction to form $\sigma^*$ bonds. However, this is a rough estimation and few thorough investigations on this part, which can be comparatively discussed with our results have been reported up to now. Further research on it will be published

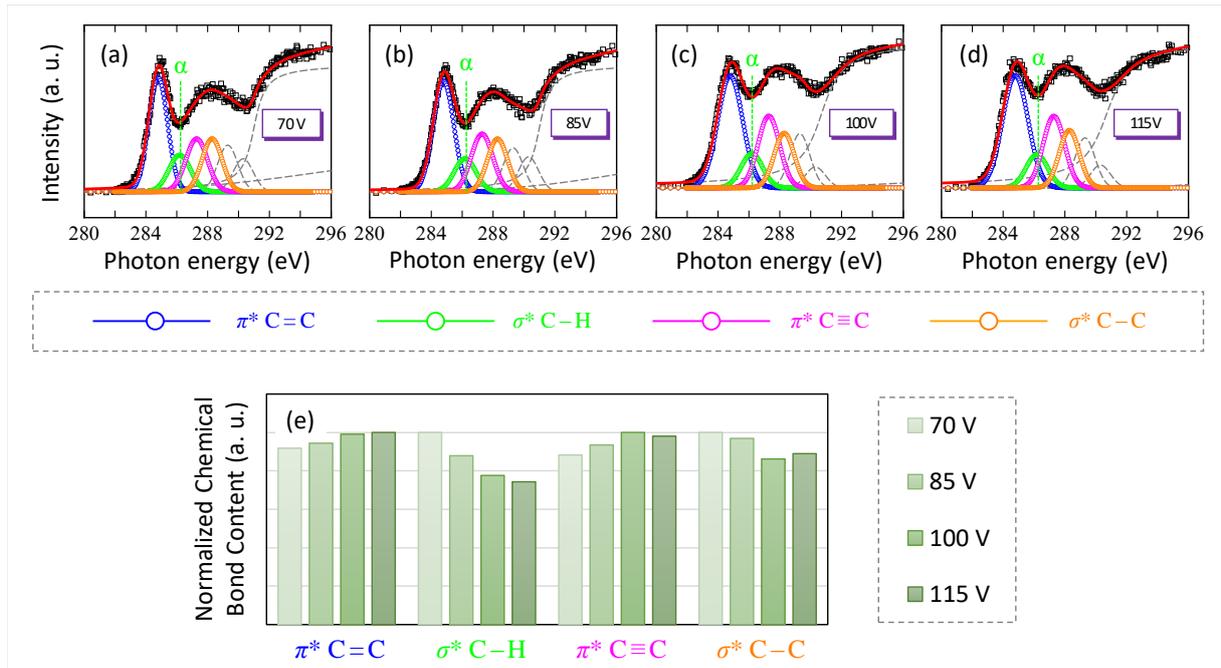

**Fig. 5.** Magnified NEXAFS C $K$-edge spectra in UNCD/a-C:H films synthesized with applied voltages of (a) 70, (b) 85, (c) 100 and (d) 115 V, respectively, which are presented in Fig. 4. Deconvoluted spectra centered at a photon energy of 284.8 eV ($\pi^*$ C=C) (blue), 286.2 eV ($\sigma^*$ C−H) (light green), 287.3 eV ($\pi^*$ C≡C) (magenta), 288.3 eV ($\sigma^*$ C−C) (orange), 289.3 eV (dashed), 290.3 eV (dashed), and 303.8 eV (dashed), are illustrated as similarly fitted with a previous work.[67] The cumulative fitting peaks are exhibited with red curves. Relative amounts of each carbon hybridization are represented in (e).

elsewhere.

Next, we study the region of less than ~291 eV in a photon energy, and this area is emphasized in Figs. 5(a)–(d). All the spectra can be approximately deconvoluted with gaussian and arctangent functions after the backgrounds are subtracted and pre- and post-edges are normalized. The fitting parameter, FWHM in the deconvoluted spectrum centered at 284.8 eV decreases from 1.95 to 1.44 eV with an increase of applied voltage. FWHM in the spectra at 286.2 and 287.3 eV is set 1.65 eV and the other components 1.41 eV, all of which fall within an acceptable error range. These value tendencies are similar to those in a literature referred to this time, which would be obtained via total electron yield (TEY) mode.[67] All of the above analytical process were performed with XAFS analysis software 'Athena'.[68]

Firstly, we make mention of the specific decomposed spectra to confirm a robustness of the following discussion. A peak originating from $\sigma^*$ C−B that will appear at 286.9 eV confirmed in the films by PLD[69] couldn't be observed since its amount is too small to be detected by NEXAFS. In addition, two additional peaks that have not been observed in UNCD/a-C:H films appear at 289.3 and 290.3 eV, and they might come from carbonyl ($\sigma^*$ C=O) and carboxy ($\sigma^*$ COOH) groups that have been detected in a-C:H films[70] and a variety of carbon-related materials such as graphene oxides,[71] respectively in the vicinities of the regions. On the occasion of the discussion afterwards, they are not considered and any remarks on them aren't done since they are susceptible to the steps denoting their ionization potentials, whose width are different each other due to the fitting issue, and are largely affected with the step tails in actual.

The prominent appearance change in the normalized NEXAFS spectra expressed with black in Figs. 5(a)–(d) starts from the right shoulder of a distinct peak at around a photon energy of 285 eV to an outset of third arch at ~290.5 eV. These areas obviously swell accompanied with an increase of an applied voltage, and they are composed of mainly four constituents originating from $\pi^*$ C=C, $\sigma^*$ C−H, $\pi^*$ C≡C, $\sigma^*$ C−C colored in Figs. 5(a)–(d), and they are studied from the following. A clear intensity transition of a peak positioned at 287.3 eV that derives from $\pi^*$ C≡C has been confirmed [see Figs. 5(a)–(d)]. This is gradually weakened in accordance with a decrease of an applied voltage, i.e. more hydrogenated, and an amount of the chemical bond is also decreasing actually as shown in Fig. 5(e). When we move an attention to a peak at 286.2 eV stemming from $\sigma^*$ C−H, it doesn't illustrate a quantitative change at a glance. Given that the peak intensity in the vicinity of the peak is stronger in higher applied voltage as mentioned earlier, the relative quantity of this bonding is varied, which is reflected well on Fig. 5(e), and it can be confirmed intuitively by focusing on label α. It leads to an indication that $\sigma^*$ C−H is formed by hydrogenation process. Likewise, more $\sigma^*$ C−C exists in more hydrogenated films, indicating an incorporation of steadier a-C:H matrices. Conversely, the content of $\pi^*$ C=C in the films diminishes in accordance with hydrogenation.

These quantitative transitions of any chemical bonds should be an evidence that the hydrogen termination certainly occurs inside the films because well-known nature of hydrogenation preferentially reacts with C≡C and C=C into C−H, which is often utilized as catalysis. These double- and triple-bonded carbons are localized at amorphous part or GBs between UNCD grains and amorphous as mentioned earlier in Sec. I. From previous studies on the electrical conductivity[34] and rectifying action[41] of p-n heterojunction diodes including p-type UNCD/a-C:H thin films, hydrogenation would be more effective on GBs rather than a-Cs, which will be supported supplementally by a previous study of MCD and NCD films synthesized by CVDs concluding minification of the diamonds accelerated an incorporation and retention of hydrogen trapped in the films,[72] in other words, our additionally hydrogenated films own the larger size and the amount of UNCD crystallites as mentioned in Sec. III B and a-C:Hs with less defects from more $\sigma^*$ C−C containing, which makes the films with less room for hydrogen incorporated into other defects except for GBs. Therefore, it is persuasive interpretation in the results obtained this time that such hydrogenation by controlling an applied voltage would terminate $\pi^*$ C≡C as well as $\pi^*$ C=C existing as dangling bonds at GBs to be transformed into $\sigma^*$ C−H. This explanation never contradicts with the results of FT–IR analysis in the previous subsection, instead it can be in good agreement with the discussion on FT–IR spectra.

### D. Correlation with optical and electrical properties

These results of hydrogenation effects on chemical bonding states of the films are of interest in possible applications toward optoelectronic devises especially for deep-UV photodetectors as described in Sec. I because these effects would be much active at GBs where the optical and electrical characteristics that the whole films possess are dominant in such composite materials with amorphous structures. We will, therefore, give detail descriptions correlated with such properties that have been investigated in the UNCD/a-C:H films synthesized by PLDs and CAPDs to date. Light absorption coefficient of materials is improved by amorphization in general since the randomness accompanied with defects such as dangling bonds gives birth to a few localized states nearby the Fermi level, which enables optical transition not via phonon but via those levels, instead[73] (see Ref. 74 concerning a theoretical study highlighting such defects at GBs in CVD UNCD films). In addition to it, several spectral response measurements have confirmed that UNCD grains surely act as a generation source of electron-hole pairs induced with UV irradiation in a couple of p-n heterojunction diodes including boron-doped UNCD/a-C:H thin films similarly fabricated by PLDs,[75,76] as it can also be indicated by rectification action of the diodes by CAPD from our recent work.[41] Our experimental results have shown more incorporated UNCD grains (Sec. III B) and stable a-C:H matrices containing less defects in the films, which is implied with an increase of the bonding state $\sigma^*$ C−C (Sec. III C) by hydrogenation. PVD UNCD/a-C:H films proud excellent optical absorptions in the UV range as compared with similarly fabricated a-C and DLC films,[77] which indicates more UNCDs as the generation source should overcome the above-mentioned demerit by diminishing the electronic states of such defects permitting photocarrier to be excited toward conduction bands and would be applicable to our films. Hence, our work has implied that the hydrogenation technique will increase the absorption coefficients of the films in the UV range from the viewpoint of chemical bonding configurations.

Here, when we review the optical bandgap of the CAPD films, three different bandgaps were detected; (i) indirect gap with its value of 1.63 eV[43] to 1.7 eV[23,57] stemming from a-Cs and a-C:Hs, (ii) direct gap of 2.9 eV[23] to 3.0 eV[57] from GBs and (iii) indirect one equivalent to more than 5.4 eV[23,57] deriving from diamond grains. We should, however, note that those values are different depending on the preparation technique, the sorts of (and the amounts of) dopants, and the preparation conditions; concretely, varying from 1.0 eV–1.7 eV and 2.2 eV–3.0 eV in (i) and (ii), respectively as far as to be reported (see Refs. 20, 43, 75, and 77 for the other cases). These different optical gaps coming from each constituent would produce multiple carrier transportation routes and mechanisms. One of the previous examinations on the temperature-dependence of electrical conductivity of the above diodes showed that these diodes fabricated under a hydrogen pressure of over a certain point (6.7 Pa) would follow three-dimensional variable-range hopping (VRH) mechanism,[35] while they also transition from semimetals to semiconductors with hydrogenation to a certain degree.[35] The hopping conduction will be highly conceivable considering their components, namely large amounts of amorphous fractions and GBs that contain numerous defects creating impurity levels between diamond bandgaps, presumably above the Fermi level supported by a molecular-dynamics simulation study,[74] and force the photo-induced carrier arising at UNCD grains to transport through them three-dimensionally. Where these carrier transportations are conducted is discussed in the following.

From the results of the current-voltage characteristics of the p-n heterojunction diodes,[41] the calculated ideality factor took the value of 2 to 3, which will indicate the carriers go through tunneling and generation-recombination (G–R) processes, and didn't show evident tendency toward the hydrogenation despite the enhancement of photoresponsivity detected from the reverse currents. The G–R process may occur via the localized levels attributed to defects and distortion of bond angle in the a-C:H matrix. Meanwhile, some research have concluded differently that the tunneling process will occur at UNCDs[78] or GBs,[32,75,76,79] however it is postulated here that both of these hypotheses are possible. Our experimental results and in-depth considerations of several spectroscopic studies have clarified the unsolved matters situated in between the structural and the optical, electrical properties. Since the enlargement of UNCD grain diameter should lessen the provability of tunneling through the grains owing to the thickened potential barriers, the tunneling via GBs is getting more dominant with increasing hydrogenation. G–R process that occurs via the localized states of amorphous would be suppressed by hydrogenation when it exerts an effect on the defects. The amount of G–R process, on the other hand, becomes secondary since the localized defect states at around the midgap and $\pi^*$ levels at conduction band edge decrease, which is revealed from generation of $\sigma^*$ C−C states and decreases of several $\pi^*$ carbon hybridizations as demonstrated in Sec. III C, and the optical gap increases with hydrogenation. From the above descriptions, it has been clarified that the GBs become the most dominant as a carrier pathway and hydrogenation terminated the dangling bonds at GBs and prevented the carrier recombination, resulting in an enhancement of photoresponsivity in the UV region. Additionally, it also demonstrated that dangling bonds in UNCD/a-C:H films can be a demerit as a role of carrier trap centers rather than a support of UV light absorptions due to the carrier excitation via such levels. To summarize, generation and transportation mechanisms of photogenerated carrier in UNCD/a-C:H films can be explained as follows. (i) Electron-hole pairs are created by irradiation of UV rays. (ii) Photogenerated carriers transport three-dimensionally with VRH process in the p-n junction diodes. (iii) Possible carrier paths are multiple; tunneling via GBs and/or UNCDs and G–R conduction in a a-C:H matrix. Photocurrent through the GBs is the most dominant and the second is via a a-C:H matrix, considering the film morphologies. Tunneling conductions in UNCD grains may be fewer compared to other two conductions. Regarding the phenomena at junction fractions, fast exudations of carrier out of Si substrates into boron-doped UNCD/a-C:H films deposited via PLD is reported,[78] however the details on it are not entirely sure and the further investigations should be required to comprehend the mechanisms completely. Our research has revealed from the viewpoint of chemical bonding states that the hydrogenation technique makes the carrier transportations through the GBs more predominant and paves the way for carriers by effectively terminating dangling bonds at GBs mainly acting as carrier trap centers.

## IV. SUMMARY, CONCLUSIONS AND OUTLOOK

We have investigated the influence of hydrogenation on UNCD/a-C:H thin films by use of several spectroscopic techniques and discussed on it in depth, and the correlations between the chemical bonding transitions revealed from the spectroscopic studies and the optical and electrical properties are also given. The summary of each spectroscopic study in Sec. III is as follows.

*XPS characterization*: XPS technique cannot detect hydrogen-terminated carbons as the same results have been reported several times in some similar materials. On the contrary, there are a couple of reports stating the opposite results up to now. This issue is under controversy at present.

*FT–IR characterization*: From the intensities of absorbance, the existence of denser embedded UNCD crystallites in more hydrogenated films has been inferred. Several deconvoluted spectra of C−H stretching bands exhibit explicit quantitative transition of $sp^2$- to $sp^3$-hybridized hydrocarbons induced by more hydrogenation. This result is an evidence of hydrogen-termination of double-bonded system at GBs between UNCD crystallites and an a-C and/or a a-C:H, and/or amorphous area.

*NEXAFS characterization*: It is demonstrated that additional hydrogenation induced by reducing an applied voltage to an arc gun denotes pronounced $\sigma^*$ features implying more stabilized UNCD crystallites and/or a a-C:H matrix, and saturates $\pi^*$ C≡C and $\pi^*$ C=C that behave as dangling bonds at GBs between UNCD crystallites and a-C and/or a-C:H fractions. Increased bonding states of $\sigma^*$ C−H and $\sigma^*$ C−C imply that more existence of hydrogen-terminated carbon atoms at GBs and stable amorphous fractions accompanied with less defects, respectively.

These comprehensive spectroscopic studies of hydrogenation effects on chemical bonding structures of the thin films opened the way for an enhancement of applicable device capability such as UV-photodetectors. The obtained results correlated with the optical and electrical characteristics present the following conclusions. (i) G–R process undergoes mostly GBs between UNCD grains and a-C and/or a-C:H matrices and becomes dominant with hydrogenation. (ii) Tunneling conduction through diamond grains is restrained due to the crystal enlargement bringing about the widen potentials. (iii) Incorporated hydrogen atoms paved a path for photo-currents running through GBs by terminating dangling bonds that captures induced-carrier. (iv) Hydrogenation stabilized amorphous structures by reducing the defects inside them and removed electrical states provably above the Fermi level that makes a detour for carriers possible to be excited to the conduction band, which might lead to an extremely large light absorption coefficients original to the UNCD/a-C:H thin films stated in Sec. I. However, a couple of advantages as photovoltaics stated in (i), (ii) and (iii) defeat this nature and it can be confirmed from a recent research on p-n heterojunction diodes including the films fabricated under totally the same conditions.[41]

This work has demonstrated how to fabricate higher-quality films for optoelectronic applications as well as its detail mechanism and would be a guide to the future research and realization of the commercialization. To be specific, this nature would be applicable not only to nitrogen-doped UNCD/a-C:H films,[36–40,42] but also to the films prepared via PLDs which deposit thin films with similarly high compositeness. Concerning the structural evaluations, Raman spectroscopic technique possessing several drawbacks in our carbon-related films with higher compositeness has recently overcome them to probe the detail comprehensive structures including the grain size of embedded diamonds directly and nondestructively,[80] which will accelerate the development of our films toward the practical device applications.

## ACKNOWLEDGEMENTS

This work is succeeded from Yoshitake Research Group (Kyushu University). The author N. N. acknowledges to Dr. Satoshi Takeichi (OSG Corporation) for his cooperation of sample fabrications, XPS and NEXAFS experiments, Dr.